\documentclass[reprint,aps,prd,onecolumn,showpacs,amsmath,nofootinbib]{revtex4-1}

\newcommand{\beqa}{\begin{eqnarray}}
\newcommand{\eeqa}{\end{eqnarray}}
\usepackage{natbib} 
\usepackage{graphicx}
\usepackage{epsfig}
\usepackage{color}

\usepackage{graphicx}
\usepackage{dcolumn}
\usepackage{bm}%

\begin{document}

\title{The contributions of dark matter annihilation to the 
global 21cm spectrum observed by the EDGES experiment}

\author{Yupeng Yang$^{1,2}$}

\affiliation{$^1$Collage of Physics and Electrical Engineering, Anyang Normal University, Anyang, 455000, China\\
$^2$Joint Center for Particle, Nuclear Physics and Cosmology, Nanjing, 210093, China
}

\begin{abstract}
The EDGES experiment has observed an absorption feature in the global 21cm spectrum with 
a surprisingly large amplitude. 
These results can be explained by decreasing the kinetic temperature of baryons, 
which could be achieved through the scattering between the baryons and cold dark matter particles. 
It seems that the mostly researched dark matter annihilation model is not able to explain such a large amplitude, 
since the interactions between the particles produced 
by the dark matter annihilation and the particles that have been present in the Universe 
could increase the baryonic temperature. Recently, C. Feng and G. Holder have suggested that
the large amplitude in the global 21cm spectrum could be produced by considering 
the possible excess of the early radio radiation. 
In this paper, we propose that the dark matter annihilation still works to explain 
the large amplitude observed by the EDGES experiment. 
By considering the possible excess of the early radio radiation,  
the large absorption amplitude in the global 21cm spectrum could be produced even 
including the dark matter annihilation. 

\end{abstract}

\maketitle

\section {Introduction} 
As an important way of exploring the "dark ages" of the Universe, 
the global 21cm spectrum has been studied in theory by many works; see, e.g., Refs.~\cite{21cm_review_2,21cm_review_3,21cm_review_1}. 
Recently, the EDGES experiment reported the observational results on 
the global 21cm spectrum, which finds an absorption feature at the redshift $z\sim 17$ 
with a surprisingly large 
amplitude $T_{21} \sim 500 $~mK~\cite{nature-1}, twice as large as expected. 
The amplitude of the global 21cm spectrum is accounted for the 
competitions among the kinetic temperature ($T_{k}$), 
the CMB thermodynamic temperature ($T_{\rm CMB}$) and 
the spin temperature ($T_{s})$. 
One possible way of explaining the observed large amplitude is decreasing 
the kinetic temperature $T_{k}$, which could be achieved if the scattering 
between the baryons and cold dark matter particles is present~\cite{nature-2,Munoz:2018pzp,Fialkov:2018xre,Barkana:2018qrx}. 
Another possible way 
is enhancing the temperature of the cosmic radio background~\cite{Fraser:2018acy,arcade-2,Ewall-Wice:2018bzf}. 
In Ref.~\cite{arcade_pri}, the ARCADE-2 experiment 
reported the excess of the cosmic radio background in the frequency $\nu \lesssim $1~GHz, 
the corresponding temperature can be fitted with a form 

\beqa
T(\nu) = T_{0}+T_{e}\left(\frac{\nu}{\rm 1~GHz}\right)^{\alpha},
\label{tcmb}
\eeqa
where $T_{0}=2.729\pm 0.004$~K is the CMB thermodynamic temperature at $z=0$, 
$T_{e}=1.19\pm 0.14$~K and $\alpha=-2.62\pm 0.04$.  
The excess of the cosmic radio background, not explained easily by 
the standard sources, 
could be from the early radio sources such as the radio-loud quasars; 
see, e.g., Refs.~\cite{2009arXiv0901.0559S,Kogut,Singal,Bolgar:2018rry,arcade-summary}. 
Some authors have found that the radio excess 
could be explained by the dark matter annihilation~\cite{arcade-3,arcade-4,arcade-5}.\footnote{One should notice that 
the radio excess explained by the dark matter annihilation is observed at $z=0$.}
The main point is that the electrons produced during the dark matter annihilation 
could emit the synchrotron radiation within the cosmic magnetic field. 
On the other hand, the evolution of the Universe could be influenced 
by the dark matter annihilation~\cite{deltaTb,binyue,j_alpha,lezhang,xlchen,prd_1,DAmico:2018sxd}. 
One of the influences is heating 
the intergalactic medium (IGM) and enhancing the kinetic temperature $T_{k}$. 
Therefore, it seems that 
the dark matter annihilation is not able to explain the observed large amplitude 
of the global 21cm spectrum~\cite{Berlin:2018sjs,Kang:2018qhi}. 
In Ref.~\cite{arcade-2}, the authors found 
that the amplitude of the 
global 21cm spectrum could reach a very large amplitude, $T_{21}\sim 1100$~mK, 
even with 10 percent of the observed radio excess. 
In this paper, we propose that although the dark matter annihilation 
can increase the kinetic temperature, it could still explain 
the observed large absorption amplitude in the global 21cm spectrum. 
By considering the possible excess of the early radio radiation 
caused by, such as the early radio sources, even the dark matter annihilation 
heats the IGM, 
a large absorption amplitude in the global 21cm spectrum could appear, 
which is consistent with the 
observational results of the EDGES experiment. 

This paper is organized as follows: In Sec. II, we review the basic quantities of the global 21cm 
signal. The influences of dark matter annihilation and the first stars on IGM are investigated in Sec. III. 
In Sec. IV, we investigate the global 21cm spectrum including the dark matter annihilation and 
the excess of the cosmic radio background. 
The conclusions and discussions are given in Sec. V.

\section{The basic quantities of the global 21cm signal}

In this section, we briefly review the basic quantities of the global 21cm signal. 
For detailed discussions, one can refer to, e.g., Refs.~\cite{21cm_review_1,21cm_review_2,21cm_review_3} 
and references therein.

The 21cm signal is accounted for by the transition of the hyperfine split of the hydrogen atoms.  
The ground state of hydrogen can split into triplet and singlet states, 
and the energy change of the two levels is $E=5.9\times 10^{-6} \rm~eV$ 
corresponding to the wavelength of photon $\lambda = 21$cm. 
The spin temperature $T_s$ is defined as 

\beqa
\frac{n_1}{n_0}=3\mathrm{exp}\left(-\frac{T_\star}{T_s}\right),
\eeqa
where $n_1$ and $n_0$ are the number densities of hydrogen atoms in triplet and 
singlet states, and $T_{\star}$ is the equivalent temperature corresponding to the transition energy.

The spin temperature is mainly effected by 
(i) the background photons; (ii) the collisions of the hydrogen atoms with 
other particles; (iii) the resonant scattering of $\rm Ly\alpha$ photons. 
Including these factors, the spin temperature can be written as~\cite{binyue,deltaTb} 

\beqa
T_{s}=\frac{T_{\rm CMB}+(y_{\alpha} + y_c)T_{k}}{1+y_{\alpha}+y_{c}},
\label{eq_ts}
\eeqa 
where $y_\alpha$ corresponds to the Wouthuysen-Field effect, and 
in this work we adopt the form used in Refs.~\cite{Kuhlen,binyue},

\beqa
y_{\alpha}=\frac{P_{10}T_{\star}}{A_{10}T_{k}}\mathrm{e}^{-0.3(1+z)^{0.5}T_{k}^{-2/3}
\left(1+0.4/T_{k}\right)^{-1}}
\eeqa
where $A_{10}=2.85\times 10^{-15}s^{-1}$ is the Einstein coefficient of the hyperfine 
spontaneous transition. $P_{10}=1.3\times 10^{9}J_{\alpha}$, is the de-excitation rate of 
the hyperfine triplet state due to Ly${\alpha}$ scattering. $J_{\alpha}$ is the intensity 
of Ly${\alpha}$ radiation~\cite{xlc21cm-1,Ciardi}, 

\beqa
J_{\alpha} = \frac{c(1+z)^2}{4\pi}\int^{z_{max}}_{z}
\frac{\epsilon(\nu^{\prime},z^{\prime})}{H(z^{\prime})}dz^{\prime}
\eeqa
where $\nu^{\prime} = \nu_{\alpha}(1+z^{\prime})/(1+z)$. 
$\epsilon(\nu^{\prime},z^{\prime})$ is the 
comoving photon emissivity~\cite{xlc21cm-1,Ciardi,Barkana:2004vb,21cm_review_1,21cm_review_2}. 
Theoretically, 
the star formation affected by dark matter annihilation could 
influence the $\rm Ly\alpha$ radiation, see e.g., Refs.~\cite{Ripamonti:2006gr,binyue,Smith:2012ng}. 
In this work, 
we neglect this effect which will be discussed detailedly in the near future work. 
In Eq.~(\ref{eq_ts}), $y_c$ corresponds to the collision effect between hydrogen atoms, electrons 
and protons, and in this work, we adopt the form used in Refs.~\cite{binyue,Kuhlen,Liszt,epjplus_2}, 

\beqa
y_{c}=\frac{\left(C_{\rm HH}+C_{\rm eH}+C_{\rm pH}\right)T_{\star}}{A_{10}T_{k}}
\label{yalpha}
\eeqa
where $C_{\rm HH,eH,pH}$ are the de-excitation rate and we adopt the forms used in Refs.~\cite{Kuhlen,Liszt}.

In general, the mostly used quantity for the observation of 
the global 21cm signal is the brightness temperature $T_{21}$ 
which can be written as~\cite{deltaTb,Ciardi} 

\beqa
T_{21} = &&26(1-x_{e})\left(\frac{\Omega_{b}h}{0.02}\right)
\left(\frac{0.3}{\Omega_{m}}\right)^{\frac{1}{2}}
\left(\frac{1+z}{10}\right)^{\frac{1}{2}} \nonumber \\
&&\times\left(1-\frac{T_{\rm CMB}}{T_{s}}\right)\rm mK,
\label{eqt21}
\eeqa
where $x_e$ is the fraction of free electrons.

\section{The influences of dark matter annihilation 
and the first stars on the intergalactic medium}

Dark matter as the main component of the Universe has been confirmed by 
many observations while its nature is still unknown. 
There are many dark matter models and the mostly researched one is 
weakly interacting massive particles (WIMPs)~\cite{dm_1,dm_2,xjb}. 
According to the theory, 
WIMPs could annihilate into normal particles, such as photons, electrons 
and positrons. There are interactions between the particles produced by 
the dark matter annihilation and the particles present in the Universe. 
These interactions could influence the evolution of the IGM and 
the main influences on IGM are heating, ionization and excitation~\cite{deltaTb,binyue,j_alpha,lezhang,xlchen,prd_1}. 
Including the dark matter annihilation, the changes of the ionization 
degree ($x_e$) and the temperature of IGM ($T_k$) with the time are~\cite{deltaTb,binyue,j_alpha,lezhang,xlchen,prd_1} 

\beqa
(1+z)\frac{dx_{e}}{dz}=\frac{1}{H(z)}\left[R_{s}(z)-I_{s}(z)-I_{\rm DM}(z)\right],
\eeqa

\beqa
(1+z)\frac{dT_{k}}{dz}=&&\frac{8\sigma_{T}a_{R}T^{4}_{\rm CMB}}{3m_{e}cH(z)}\frac{x_{e}}{1+f_{\rm He}+x_{e}}
(T_{k}-T_{\rm CMB}) \\ \nonumber
&&-\frac{2}{3k_{B}H(z)}\frac{K_{\rm DM}}{1+f_{\rm He}+x_{e}}+T_{k},
\eeqa
where $R_{s}(z)$ and $I_{s}(z)$ are the standard recombination rate and ionization rate, respectively. 
$I_{\rm DM}$ and $K_{\rm DM}$ are the ionization rate and heating rate caused by the 
dark matter annihilation~\cite{xlchen,prd_1,lezhang}. 
For our purposes, the influences of dark matter annihilation on the 
evolution of the IGM should be included in order to 
investigate the changes of $T_{k}$ with time. In this paper, we follow 
the methods presented in Refs.~\cite{lezhang,xlchen,prd_1} and 
modify the public code RECFAST~\footnote{http://camb.info/} 
to include the effects of dark matter annihilation. 
Including the dark matter annihilation, for example, 
at the redshift $z\sim 20$, the kinetic temperature $T_{k}$ and ionization degree $x_e$ 
could reach up to $T_k\sim 100$~K 
and $x_e\sim 0.001$, respectively~\cite{deltaTb,valdes,binyue}.

If we do not include the dark matter annihilation, there are several standard processes that 
could influence the evolution of IGM~\cite{Furlanetto:2006tf,21cm_review_2,21cm_review_3,21cm_review_1}. 
At high redshift, Compton scattering between CMB photons and the free electrons 
is the main source of heating. After the formation of the first luminous 
structures, X-rays from e.g. galaxies and quasars are dominant for heating. 
The luminosity of X-ray is proportional to the star formation rate, 
which is proportional to the differential increase
of the baryon collapse fraction~\cite{21cm_review_3,21cm_review_2,21cm_review_1,Furlanetto:2006tf}. 
The energy deposited in the IGM from X-rays can be written as~\cite{Furlanetto:2006tf,binyue} 

\beqa
\epsilon_{X}(z) = &&1.09\times 10^{-31}f_{X}f_{\star}\left[\frac{\rho_{b,0}(1+z)^3}
{\rm M_{\odot}Mpc^{-3}}\right]\left|\frac{df_{\rm coll}(z)}{dt}\right|, 
\label{epsilon}
\eeqa 
where $f_{\rm coll}(z)$ is the collapse fraction~\cite{21cm_review_1,Furlanetto:2006tf}. 
$f_X$ is a correction factor referring to the differences 
of the X-rays between the low and high redshifts. 
Given the fact that there are a lot of uncertainties 
for the X-rays from the high redshift objects, $f_X$ is model dependent and  
in general $f_{X} \gtrsim 1$~\cite{Furlanetto:2006tf}.
\footnote{For some models, $f_X$ could be smaller than unit, $f_{X} \sim 0.2$~\cite{Furlanetto:2006tf,Glover:2002rf}.}
In this work, we take the conservative and reasonable value as $f_{X}= 1$. 
$f_{\star}$ is the star formation efficiency and is model dependent. 
In Ref.~\cite{Wada:2007wc}, the authors found that 
the star formation efficiency is $f_{\star} \sim 0.001-0.01$ for 
normal spiral galaxies and $f_{\star}\sim 0.01-0.1$ 
for starburst galaxies, respectively. 
In this work, we take the conservative value as 
$f_{\star}=0.001$~\cite{Gnedin:1999fa,Furlanetto:2006tf,Wada:2007wc}. 
The intensity 
of $\rm Ly{\alpha}$ radiation from the X-rays can be written as~\cite{Chen:2006zr,21cm_review_2} 

\beqa
J_{\alpha,X}= \frac{c}{4\pi}\frac{1}{H(z)\nu_{\alpha}}
\frac{\epsilon_{X}}{h\nu_{\alpha}}.
\eeqa

The scattering between the neutral hydrogen atoms and 
the photons in the Lyman-series resonances could also heat the IGM. 
In Refs.~\cite{xlc21cm-1,Rybicki:2006th}, the authors found that 
the energy deposition rate of this process is very small and 
we neglect this process in this work. Another heating source 
is the shocks in the IGM and the shock heating is also model 
dependent. In Ref.~\cite{21cm_review_1}, the authors found that 
the effects of shock heating on the IGM are $\lesssim 10\%$. 
In this work, we do not include this process.
 
\begin{figure}
\epsfig{file=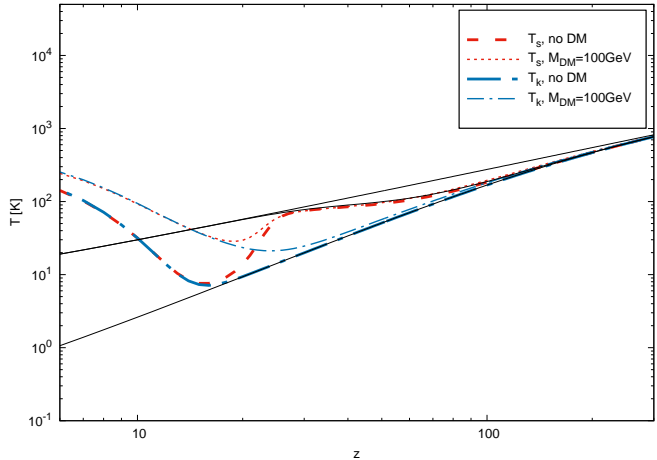,width=0.5\textwidth}
\caption{The evolution of temperature $T_s$ and 
$T_k$ for the cases with and without dark matter. 
For comparison, the evolution of $T_s$, $T_k$ and $T_{\rm CMB}$ 
for the case without reionization sources is also shown (thin solid black lines). 
Here we set the parameters of dark matter 
as $M_{\rm DM} = 100$ GeV and $\rm \left<\sigma v\right>=3\times 10^{-26}cm^{-3}s^{-1}$.}
\label{tks}
\end{figure}

\begin{figure}
\epsfig{file=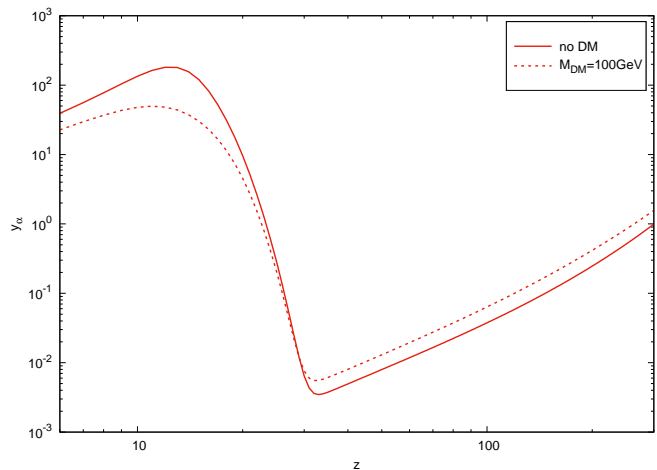,width=0.5\textwidth}
\caption{The evolution of $y_{\alpha}$ 
for the cases with and without dark matter. Here we set the parameters of dark matter 
as $M_{\rm DM} = 100$ GeV and $\rm \left<\sigma v\right>=3\times 10^{-26}cm^{-3}s^{-1}$.}
\label{yalphafig}
\end{figure}

The evolution of $T_s$ and $T_k$ 
with and without dark matter annihilation is shown in Fig.~\ref{tks}. 
For comparison, the evolution of $T_s$, $T_k$ and $T_{\rm CMB}$ 
without the influences of reionization sources is also shown (thin solid black lines). 
Compared with the case without reionization sources, 
the temperature of IGM increases after the redshift $z\sim 20$ 
due to the presence of the heating sources. The spin temperature $T_s$ decouples 
from $T_k$ at the redshift $z\sim 200$ and is coupled to $T_k$ again after the redshift $z\sim 20$. 
The evolution of $y_{\alpha}$ with time is shown in Fig.~\ref{yalphafig}. 
One of the factors that could influence $y_{\alpha}$ is the 
intensity of the $\rm Ly{\alpha}$ radiation. 
In Ref.~\cite{binyue}, the authors investigated the evolution 
of $J_{\alpha}$ for the cases 
with and without dark matter annihilation (Fig.~5 in Ref.~\cite{binyue}). 
It was found that 
the $\rm Ly{\alpha}$ background is mainly from the dark matter annihilation during the dark ages, 
while the contributions from the first stars are dominant after the redshift $z\sim 30$. 
In Fig.~\ref{yalphafig}, 
it can be seen that, due to the dark matter annihilation, the values of $y_{\alpha}$ 
are larger than that of without dark matter at early times, while 
the strong gas heating effect reduces the values of $y_{\alpha}$ 
after the redshift $z\sim 30$. 
One should notice that the evolution of $T_s$ and $T_k$ 
is model dependent, and for detailed discussions, one can 
refer to e.g. Refs.~\cite{21cm_review_1,deltaTb,Furlanetto:2006tf}.

\section{The global 21cm spectrum including the excess of the cosmic radio background}

As shown in the above section, the temperature of IGM increases due to the dark matter annihilation. 
Therefore, the observed large amplitude in the global 
21cm spectrum could not be explained 
easily if the dark matter annihilation were included. 
In this section, we show that the observed large amplitude 
in the global 21cm spectrum could still be explained even including the dark matter annihilation 
if the excess of the cosmic radio background is included. 

As mentioned in Sec. I, the excess of the cosmic radio background in the frequency $\nu \lesssim $1~GHz 
has been observed by the ARCADE-2 experiment. 
The excess could not be explained easily by the standard sources, 
such as the galactic emission or 
extragalactic sources counts~\cite{arcade-summary,Fornengo:2014mna}. 
In Ref.~\cite{Fang:2015dga}, the authors found that the 
radio excess 
could be explained in the presence of the magnetic turbulence and shocks 
in merging galaxy clusters, where the non-thermal electrons 
are re-accelerated via $\rm Alfv\acute{e}n$ waves. 
Other possible sources of the radio excess 
would be from the high redshift objects. 
Considering the uncertainties 
of the radio sources in early times, the excess fraction of the radio background 
at the high redshift 
would be small. 
Moreover, at the high redshift, including the radio excess 
the intensity of the radio radiation background 
would be larger than that of CMB at a rest wavelength of 21cm~\cite{arcade-2}. 
Therefore, following Ref.~\cite{arcade-2}, 
we write the corresponding temperature of the radio 
radiation background as

\beqa
T_{\rm CMB}(\nu) = T_{0}+\beta T_{e}\left(\frac{\nu}{\rm 1~GHz}\right)^{\alpha},
\label{tcmb2}
\eeqa
where $\beta$ is a free parameter describing the excess fraction of the cosmic radio background at early times. 
For our purposes, we set 
$\nu=1420\rm MHz/(1+z)$. It should be noticed that one should use the form 
$T_{\rm CMB}=T_{\rm CMB}(\nu)(1+z)$ to calculate 
the brightness temperature $T_{21}$ in Eq.~(\ref{eqt21}).

\begin{figure}
\epsfig{file=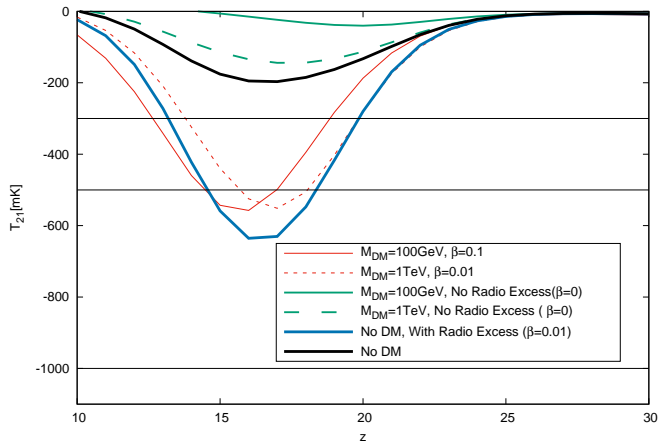,width=0.5\textwidth}
\caption{The global 21cm spectrum in the redshift $10\leq z\leq 30$ 
including the dark matter annihilation and the excess of the cosmic 
radio background. 
Here we set the thermally averaged cross section of dark matter annihilation 
as $\rm \left<\sigma v\right>=3\times 10^{-26}cm^{-3}s^{-1}$, 
the mass of dark matter particle 
and the parameter $\beta$ as 
$M_{\rm DM}=100~\rm GeV, \beta=0.1$ (solid red thin line); 
$M_{\rm DM}=1~\rm TeV, \beta=0.01$ (dotted red thin line). 
For comparison, we also show the global 21cm spectrum for (i) the case without dark matter 
(solid black bold line); (ii) the case without the excess of the cosmic 
radio background ($\beta=0$), 
$M_{\rm DM}=100~\rm GeV$ (solid green bold line) 
and $M_{\rm DM}=1~\rm TeV$ (dashed green bold line); 
(iii) the case without dark matter but with the excess of the cosmic 
radio background ($\beta=0.01$, solid blue bold line). 
The horizontal lines correspond to the temperature 
of the global 21cm spectrum observed by EDGES experiments, 
$T_{21}=-500^{+200}_{-500}$~mK~\cite{nature-1,DAmico:2018sxd}.
}  
\label{t21}
\end{figure}

The global 21cm spectrum 
in the redshift $10 \lesssim z \lesssim 30$ 
is shown in Fig.~\ref{t21}. For our calculations, we set the 
thermally averaged cross section of dark matter annihilation as 
$\rm \left<\sigma v\right>=3\times 10^{-26}cm^{-3}s^{-1}$. 
As shown in Fig.~\ref{tks}, the kinetic temperature $T_k$ 
is enhanced in the presence of dark matter annihilation 
after the redshift $z \sim 30$. 
For the coupling factor $y_{\alpha}$, it is depressed for the case of dark matter 
annihilation after the redshift $z\sim 30$. 
Therefore, the large absorption feature in the 
global 21 cm spectrum could not be explained easily if the dark matter annihilation were included.
From Fig.~\ref{t21}, it can be seen that including the excess of the cosmic radio background, 
the large absorption amplitude could appear in the presence of dark matter annihilation. 
The absorption amplitude of the global 21cm spectrum could reach up to 
$T_{\rm 21}\sim 550$~mK at the redshift $z\sim 16$ for $M_{\rm DM} =100$~GeV and $\beta=0.1$ (solid red thin line). 
For $M_{\rm DM}=1$~TeV and $\beta=0.01$ (dotted red thin line), 
the comparable absorption amplitude 
of the global 21cm signal appears at the redshift $z\sim 17$. 
For comparison, we also show the global 21cm spectrum for (i) the case without dark matter 
(solid black bold line); (ii) the case without dark matter but with the radio excess 
(solid blue line); (iii) the case without the radio excess but with dark matter annihilation 
(solid and dashed green bold lines). 
It can be seen that the amplitude of the global 21cm spectrum decreases 
for the case with dark matter annihilation. 
Similar effects could also be 
found e.g. in Refs.~\cite{epjplus_2,binyue,deltaTb}. 
As mentioned in Sec. II, 
shocks in the IGM could also be a heating source. The amplitude 
of the global 21cm signal would be decreased about $\lesssim 10\%$ 
if shocking heating is included. 

One issue that should be noticed is that 
the dark matter particle with a mass $M_{\rm DM}\sim 1$~TeV could also be 
used to explain the excess of the positrons flux observed by 
the DAMPE or AMS-2 experiments~\cite{Ambrosi:2017wek,leifeng-1,leifeng-2,leifeng-3}. 
On the other hand, the dark matter annihilation has influences 
on the anisotropy of the cosmic microwave background~\cite{xlchen,lezhang,prd_2015}, 
and the main influences are on the thermal and reionization history of the IGM. 
Therefore, the constraints on the parameters of dark matter could be obtained from 
the observational results, e.g. the Planck data. In Ref.~\cite{prd_2015}, the authors have got the constraints 
on the dark matter parameters as $M_{\rm DM} \gtrsim 20$~GeV for 
$\rm \left<\sigma v\right>=3\times 10^{-26}cm^{-3}s^{-1}$. 
Therefore, the parameters of the dark matter used here are within the current allowed 
parameter space. 

Another issue that should be noticed is 
that the electrons and positrons from the dark matter annihilation 
could emit the synchrotron radiation in the magnetic field of the Universe. 
This radiation could contribute to the excess of the cosmic 
radio background in the frequency $\nu \lesssim 1 \rm GHz$, 
which has been observed by the ARCADE-2 experiment~\cite{arcade,arcade-3,arcade-4,arcade-5}. 
In Refs.~\cite{arcade-3,arcade-4}, the authors found that the excess of the cosmic radio background 
could be explained by the dark matter annihilation, for example, with the dark matter mass 
$M_{\rm DM}\sim 20$~GeV and the thermally averaged cross section 
$\left<\sigma v\right> \sim 3\times 10^{-26} \rm cm^{-3}s^{-1}$ 
for the $\mu^{+}\mu^{-}$ channel. At early times, 
the contributions of the dark matter annihilation to the cosmic radio background should be 
smaller compared to the standard sources such as the radio-loud quasars~\cite{Bolgar:2018rry}, 
since the cosmic magnetic field is very weak, 
$B\lesssim 1\rm nG$.\footnote{This value is much smaller than that of the present, $B\sim 1\rm \mu G$.} 

In brief, we have considered the popular dark matter annihilation model. 
The dark matter annihilation has influences on the evolution of the Universe. 
One of the influences is heating the IGM. 
Therefore, the absorption amplitude in the global 21cm spectrum 
could be reduced or washed out if the dark matter annihilation were included. To explain 
the observational results of the EDGES experiment, one of the 
methods is enhancing the cosmic radio background. 
By considering the excess of the cosmic radio background 
at high redshift, the heating effects of dark matter annihilation could be weakened. 
On the other hand, if there is a large radio excess at high redshift, 
the absorption amplitude in the global 21cm spectrum could be 
very large. For this case, the dark matter annihilation could provide a kind of way 
of pulling the amplitude back.\footnote{In general, the dark matter annihilation can be neglected, 
and only the radio excess is included. The large absorption amplitude 
in the global 21cm spectrum could also be produced~\cite{arcade-2}. 
However, because there is no observational evidence that the 
dark matter can not annihilate, it is interesting and 
worth it to include the dark matter annihilation during 
the evolution of the Universe.

}

\section {Conclusion and discussion}
Recently, the EDGES experiment has reported a large absorption amplitude in the 
global 21cm spectrum. One possible way of explaining the results is 
decreasing the kinetic temperature. Therefore, it seems that 
the observed results could not be explained easily if 
the dark matter annihilation is included due to its heating effects on the IGM. 
In this work, we have proposed that by considering 
the excess of the cosmic radio background at early times, although the dark matter 
annihilation could increase the kinetic temperature, 
the large absorption amplitude in the global 21cm spectrum 
could also be produced. For example, 
for dark matter mass $M_{\rm DM} = 1$~TeV and $\sim 1\%$ excess of the cosmic 
radio background, the absorption amplitude of the global 21cm spectrum 
could reach up to $T_{21}\sim 550$~mK at the redshift $z\sim 17$.
 
The excess of the cosmic radio background reported by the ARCADE-2 experiment 
can not be explained easily by the standard sources. 
The radio excess would be contributed (all or partly) 
by the redshifed radiation produced at high redshift. 
Since the dark matter could annihilate into electrons, 
therefore, it is naturally expected that the synchrotron radiation 
from these electrons could contribute to the excess of the cosmic radio background. 
In Refs.~\cite{arcade-3,arcade-4}, authors found that 
the radio excess can be explained by the dark matter annihilation. 
However, one should notice that the contributions of the radio radiation 
from the dark matter annihilation are mainly from the late times 
when the intensity of the cosmic magnetic field 
is large~\cite{arcade-4}. 
At high redshift the contributions of dark matter annihilation to the
cosmic radio background would be limited for the standard dark matter
halos, because the cosmic magnetic field is weak. However, other
astrophysical sources such as the radio-loud quasars could be the
sources of the cosmic radio background excess.

In conclusion, we have shown that the popular dark matter 
annihilation model still works to explain the surprisingly large absorption amplitude 
of the global 21cm spectrum in the presence of the cosmic radio background excess, 
which could be caused by astrophysical sources such as the radio-loud quasars. 
There are some other different effects 
that could influence our final results, 
such as the effects of different dark matter annihilation models on the structure formation, 
the evolution of IGM and the $\rm Ly\alpha$ radiation. 
In theory, the constraints on the dark matter model, such as the 
lower limit constraints on the dark matter mass, could be obtained 
from the observational results of the EDGES experiment. 
Moreover, for the small dark matter mass, even including $100\%$ 
of the cosmic radio background excess it is still not possible to explain the 
large absorption amplitude in the global 21cm spectrum, 
because the heating effect from dark matter annihilation is too strong. 
More detailed calculations and 
possible constraints on the different dark matter models will be given in the 
near future work.

\section {Acknowledgements}
We thank Dr. Bin Yue, Xiaoyuan Huang, Lei Feng, Qiang Yuan and Prof. Yizhong Fan 
for very useful discussions and comments. 
We also thank the anonymous referees for the very useful comments and suggestions. 
This work is supported in part by the National Natural Science Foundation of China 
(under Grants No.11505005 and No.U1404114).
\

\bibliographystyle{apsrev4-1}
\bibliography{refs}

\end{document}